\documentstyle[color,twocolumn,prl,aps,epsf]{revtex}
\tighten

\newcommand{\lt}{\left}
\newcommand{\rt}{\right}
\newcommand{\no}{\nonumber}
\newcommand{\nn}{\nonumber\\}
\newcommand{\ds}{\displaystyle}
\newcommand{\e}{\epsilon}
\newcommand{\eq}[1]{(\ref{#1})}
\newcommand{\imag}{{\rm Im}\,}

\newcommand{\ea}{{\it et al.}}
\newcommand{\mrm}[1]{\mbox{\rm #1}}
\newcommand{\rfn}[1]{(\ref{#1})}

\newcommand{\np}[1]{{ Nucl. Phys. }{\bf #1}}

\newcommand{\pr}[1]{{ Phys. Rev. }{\bf #1}}

\newcommand{\mysection}[1]{{\bf #1}.}
\newcommand{\mysubsection}[1]{{\it #1}.}
\newcommand{\out}[1]{}

\begin{document}

\preprint{ TUM-HEP-384-00 \\ UCRHEP-T287} \draft
\twocolumn[\hsize\textwidth\columnwidth\hsize\csname@twocolumnfalse\endcsname
\title{\vspace{-3ex}\small
  FERMILAB-Pub-01/213-T,~~~~~~~~~~TUM-HEP-384-01,~~~~~~~~~~CERN--TH/2001-181,
 \hfill hep-ph/0107121\\[1mm]
\Large \bf Higgs Sector of the Minimal Left-Right Symmetric
  Model}
\author{ Gabriela  Barenboim$^1$, Martin Gorbahn$^2$, Ulrich
  Nierste$^3$ and Martti Raidal$^{3,4}$}
\address{ $^1$ Fermi National Accelerator Laboratory, 
  Batavia, IL 60510-500, USA\\
  $^2$ Physik-Department, Technische Universit\"at M\"unchen,
  D-85748 Garching, Germany\\
  $^3$ CERN--Theory, CH-1211 Geneva 23, Switzerland\\ 
  $^4$ National Institute of Chemical Physics and Biophysics, 
       Tallinn 10143, Estonia\\ 
} 
\maketitle
\begin{abstract}
  We perform an exhaustive analysis of the most general Higgs sector
  of the minimal left-right symmetric model (MLRM).  We find that the
  CP properties of the vacuum state are connected to the Higgs
  spectrum: if CP is broken spontaneously, the MLRM does not approach
  the Standard Model in the limit of a decoupling left-right symmetry
  breaking scale. Depending on the size of the CP phases scenarios
  with extra non-decoupling flavor-violating doublet Higgses or very
  light SU(2) triplet Higgses emerge, both of which are ruled out by
  phenomenology. For zero CP phases the non-standard Higgses decouple
  only if a very unnatural fine-tuning condition is fulfilled. 
  We also discuss generalizations to a non-minimal Higgs sector.
\end{abstract}
 \vglue-.2truecm]
\narrowtext
%
%
\mysection{Introduction}
Left-right symmetric models are extensions of the Standard Model (SM) based
on the gauge group SU(2)${}_{\rm R} \times\,$SU(2)${}_{\rm
  L}\times\,$U(1)${}_{\rm B-L}$ \cite{lr}. The right-handed fermion
fields are SU(2)${}_{\rm R}$ doublets and parity P is an exact
symmetry of the Lagrangian. At a high scale $v_R$ well above the
electroweak breaking scale SU(2)${}_{\rm R} \times\,$SU(2)$_{\rm L}
\times\,$U(1)$_{\rm B-L} \times\,$P is spontaneously broken to the
SM gauge group SU(2)$_{\rm L} \times\,$U(1)$_{\rm Y}$.
The U(1) charges have a physical interpretation as the difference
B$-$L of baryon and lepton number. The hypercharge Y, which is an
ad-hoc quantum number of the SM, emerges as the
combination Y$={\rm T}_{3,\rm R}+({\rm B-L})/2$, where ${\rm T}_{3,\rm
  R}$ is the third component of the right-handed isospin.  The Higgs
sector of the MLRM consists of two Higgs triplets $\Delta_R$,
$\Delta_L$ and a bidoublet $\Phi$.  The neutral component of
$\Delta_R$ acquires a vacuum expectation value (VEV) $v_R$, which
breaks the SU(2)${}_{\rm R}$ and P symmetries. The bidoublet $\Phi$
breaks the electroweak symmetry down to $U(1)_{em}$. The choice of
Higgs triplets $\Delta_{L,R}$ rather than doublets permits a Majorana
mass term and thereby small neutrino masses via the see-saw mechanism
\cite{seesaw}.  This requires $v_R$ to be very high, typically of
order $10^{10}-10^{15}~$GeV.

\mysection{Spontaneous CP violation}
Models with explicit CP violation suffer from a general problem: a
CP-non-invariant Lagrangian usually contains too many sources of CP
violation.  In most extensions of the SM, especially in the MSSM, this
problem becomes very severe: some CP-violating phases must be
fine-tuned to comply with the observed smallness of CP violating
observables.  Models with spontaneous CP violation (SCPV) are
therefore an attractive alternative, because their only few sources of
CP violation are the complex phases of the VEVs of Higgs fields.  The
MLRM with SCPV has recently attracted new attention in the context of
CP phenomenology probed in current experiments \cite{cpwrong,bf}.  The
MLRM with SCPV has the attractive feature that C,P and T are all exact
symmetries of the Lagrangian.  Both the MLRM with and without SCPV
have been studied extensively.  Yet its complicated Higgs sector has
never been analyzed completely. In \cite{cpwrong,bf,ckmcp} the case of
sizable CP-violating phases was considered aiming at the
explanation of the observed CKM CP phase through SCPV.  In
\cite{deshpande,lrh}, however, large CP-violating phases of the VEVs
have been discarded because of fine-tuning arguments. We will clarify
this point in the following.

\mysection{Model}
The SU(2)${}_{\rm L} \times\,$SU(2)${}_{\rm R} \times\,$U(1)${}_{\rm
  B-L}$ charge assignments for
the quark and lepton multiplets are $Q_L (1/2,0,1/3),$ $Q_R
(0,1/2,1/3),$ $L_L (1/2,0,-1)$ and $L_R (0,1/2,-1)$. The Higgs
multiplets are
\begin{eqnarray}
\Phi =  
\lt( \begin{array}{cc} \ds \phi_1^0 & \ds \phi_1^+ \\
                       \ds \phi_2^- & \ds \phi_2^0 \end{array} \rt),  
&\quad& 
\Delta_{L,R} =   
\lt( \begin{array}{cc}  \ds \delta_{L,R}^+/\sqrt{2} & 
                    \ds \delta_{L,R}^{++}       \\ 
                    \ds \delta_{L,R}^0 & 
                    \ds -\delta_{L,R}^+/\sqrt{2} \end{array} \rt).
\no
\end{eqnarray}
They transform under U${}_{L,R}\in\,$SU(2)${}_{\rm L,R}$ as 
$\Phi \rightarrow U_L \Phi U_R^\dagger$,
$\Delta_{L} \rightarrow U_{L} \Delta_{L} U_{L}^\dagger$ and 
$\Delta_{R} \rightarrow U_{R} \Delta_{R} U_{R}^\dagger$.
The Higgs fields transform under parity as 
$\Delta_L \leftrightarrow \Delta_R$ and  
$\phi \leftrightarrow \phi^\dagger$.
Their charge conjugation transformation reads:
\begin{eqnarray}
{\rm C}_\phi:~~\phi \leftrightarrow \tilde\phi = \tau_2 \phi^* \tau_2,
\quad
{\rm C}_\Delta:~~\Delta \leftrightarrow \tilde\Delta 
= \tau_2 \Delta^{\! *} \tau_2.
\label{c}
\end{eqnarray}
The most general C$\times$P-invariant Higgs potential is 
\cite{deshpande}
\begin{eqnarray}
\lefteqn{V(\Delta_R, \Delta_L,\Phi)  \;= \;   
-\mu_1^2  \mrm{Tr}(\phi^\dagger \phi )
-\mu_2^2 \left[ \mrm{Tr}(\tilde{\phi} \phi^\dagger ) { +}
\mrm{Tr}(\tilde{\phi}^\dagger \phi ) \right]  }\nn
&&
-\mu_3^2 \left[ \mrm{Tr}(\Delta_L \Delta_L^\dagger ) +
\mrm{Tr}(\Delta_R \Delta_R^\dagger ) \right] 
+ \lambda_1  \left[ \mrm{Tr}(\phi \phi^\dagger ) \right]^2  
\nn
&& 
+\lambda_2 \left\{ \left[\mrm{Tr}(\tilde{\phi} \phi^\dagger ) \right]^2 +
\left[ \mrm{Tr}(\tilde{\phi}^\dagger \phi ) \right]^2 \right\}  + \!
\lambda_3  \left[\mrm{Tr}(\tilde{\phi} \phi^\dagger )
\mrm{Tr}(\tilde{\phi}^\dagger  \phi)\right]   
\nn
&&
+\lambda_4 \left\{\mrm{Tr}(\phi\phi^\dagger)
\left[ \mrm{Tr}(\tilde{\phi} \phi^\dagger ) +
\mrm{Tr}(\tilde{\phi}^\dagger \phi ) \right]  \right\}  \nn
&& + \rho_1 \left\{
        \left[\mrm{Tr}(\Delta_L \Delta_L^\dagger ) \right]^2
      + \left[\mrm{Tr}(\Delta_R \Delta_R^\dagger ) \right]^2 \right\} \nn
&&+ \rho_2  \left[\mrm{Tr}(\Delta_L \Delta_L )
               \mrm{Tr}(\Delta_L^\dagger \Delta_L^\dagger ) 
   +\mrm{Tr}(\Delta_R \Delta_R )
    \mrm{Tr}(\Delta_R^\dagger \Delta_R^\dagger ) \right]  \nn
&& +\rho_3  \left[ \mrm{Tr}(\Delta_L \Delta_L^\dagger )
\mrm{Tr}(\Delta_R \Delta_R^\dagger ) \right] 
\nn
&&
+\rho_4 \left[ \mrm{Tr}(\Delta_L \Delta_L )
\mrm{Tr}(\Delta_R^\dagger \Delta_R^\dagger ) +
\mrm{Tr}(\Delta_L^\dagger \Delta_L^\dagger ) 
\mrm{Tr}(\Delta_R \Delta_R )  \right]
\nn
&&
+\alpha_1  \left\{ \mrm{Tr}( \phi\phi^\dagger )
\left[ \mrm{Tr}(\Delta_L \Delta_L^\dagger ) +
\mrm{Tr}(\Delta_R \Delta_R^\dagger ) \right]  \right\}
\nn
&&  
+\alpha_2 
  \left[ 
    \mrm{Tr}( \phi\tilde{\phi}^\dagger ) +
    \mrm{Tr}(\phi^\dagger\tilde{\phi}  ) \right]
  \left[
    \mrm{Tr}(\Delta_R \Delta_R^\dagger) +
    \mrm{Tr}(\Delta_L \Delta_L^\dagger ) \right] \nn
\out{
+\alpha_2 \left[ \mrm{Tr}( \phi\tilde{\phi}^\dagger )
 \mrm{Tr}(\Delta_R \Delta_R^\dagger)
+ \mrm{Tr}(\phi^\dagger\tilde{\phi}  ) 
\mrm{Tr}(\Delta_L \Delta_L^\dagger ) \right.
\nn
&&  
\left. \phantom{\alpha_2}
+\mrm{Tr}(\phi^\dagger\tilde{\phi}  ) 
\mrm{Tr}(\Delta_R \Delta_R^\dagger ) +
\mrm{Tr}(\tilde{\phi}^\dagger \phi ) 
\mrm{Tr}(\Delta_L \Delta_L^\dagger )  \right] 
\nn}
&& 
+\alpha_3 \left[ \mrm{Tr}(\phi {\phi}^\dagger \Delta_L \Delta_L^\dagger)
+ \mrm{Tr}( \phi^\dagger\phi \Delta_R \Delta_R^\dagger )\right]
\nn 
&& 
+\beta_1 \left[ \mrm{Tr}(\phi \Delta_R \phi^\dagger  \Delta_L^\dagger ) +
\mrm{Tr} (\phi^\dagger \Delta_L  \phi \Delta_R^\dagger ) \right]
\nn
&& 
+\beta_2 \left[ 
\mrm{Tr}(\tilde{\phi} \Delta_R \phi^\dagger  \Delta_L^\dagger) +  
\mrm{Tr}(\tilde{\phi}^\dagger \Delta_L  \phi \Delta_R^\dagger )\right]
\nn 
&& 
+ \beta_3
\left[\mrm{Tr}(\phi \Delta_R \tilde{\phi}^\dagger  \Delta_L^\dagger ) +
\mrm{Tr}(\phi^\dagger \Delta_L  \tilde{\phi} \Delta_R^\dagger )
\right]
\, . 
\label{V}
\end{eqnarray}
Here all coefficients are real. 
\out{ We discuss $\alpha_2 \not \in {\rm I \hspace{-.7mm} R}$, the
  case of a CP non-invariant Higgs potential, later.}
We discuss the case of a CP non-invariant Higgs potential later.  The
coefficients in \eq{V} must be such that $V$ has a nontrivial minimum
which leaves $U(1)_{em}$ unbroken. By using the broken symmetries one
can arrange the VEVs such as \cite{lr,deshpande}
\begin{eqnarray}
  \langle \phi_1^0\rangle=\frac{k_1}{\sqrt{2}} ,\,
  \langle \phi_2^0\rangle=\frac{k_2}{\sqrt{2}} e^{i\alpha} ,\,
  \langle \delta_L^0\rangle=\frac{v_L}{\sqrt{2}}  e^{-i\theta},\,
  \langle \delta_R^0\rangle=\frac{v_R}{\sqrt{2}},
  \no
\end{eqnarray}
with real and positive $v_{L,R}$ and $k_{1,2}$.  If
$\beta_i$,$\rho_i={\cal O}(1)$, the condition $v_R \gg k_{1,2}$
automatically enforces $k_1 k_2/(v_L v_R)$ to be of order 1. This VEV
see-saw mechanism \cite{deshpande} suppresses $v_L$, as needed to
comply with the data on the electroweak $\rho$ parameter.  
The Yukawa Lagrangian reads 
\begin{eqnarray}
-L_Y= \bar Q_L\,\hat{F}\,\phi\,Q_R + \bar Q_L\,\hat{G}\,\tilde\phi\,Q_R 
+ h.c.\,.  \label{yuk}
\end{eqnarray}
Here $\hat{F}$ and $\hat{G}$ are $3\times 3$ matrices in flavor space.
If $L_Y$ conserves CP, one can choose them real and symmetric.  Since
$\phi_1^0$ and $\phi_2^0$ couple to both up and down quarks there will
be flavor-changing neutral couplings. Here it is useful to define
\cite{ckmcp}
\begin{eqnarray}
  \left(\begin{array}{c}
      \phi^0_{-} \\ \phi_{+}^0 
    \end{array}\right)
  &=&
  \left(\begin{array}{cc}
      \cos\beta & \sin\beta\,e^{i\alpha} \\
      -\sin\beta\,e^{-i\alpha} & \cos\beta 
    \end{array}\right)
  \left(\begin{array}{c}
      \phi^0_{1} \\ \phi_{2}^{0*} 
    \end{array}\right)\,, \label{defpm}
\end{eqnarray}
with $\tan \beta=k_2/k_1$.  Analogously we define
$(\phi^+_{-}\,,\phi_{+}^+)$.  The imaginary part of the
flavor-conserving field $\phi^0_{-}$ becomes a component of the
Goldstone bosons eaten by $Z$'s. The orthogonal combination
$\phi^0_{+}$ has flavor-changing neutral couplings. For example the
flavor-changing couplings to the down quarks are given by
\begin{equation}
  {\mathcal L}^{FC\phi^0d} = \sqrt{2} \frac{k_+}{k_-^2} \phi_+^{0*} \bar{D}_L 
  V_L^\dagger M_u V_R D_R,
  \label{nondiaghiggs}
\end{equation}
where $k_\pm^2= k_1^2\pm k_2^2$ with the electroweak breaking scale
$k_+\simeq 246\,$GeV.  P and CP invariance imply $\left| V_L \right| =
\left| V_R \right|$ and calculable phases for the left and
right-handed CKM matrices. If P and CP are broken spontaneously, the
masses of the flavor-violating Higgses must exceed 10 TeV from
phenomenology \cite{bf}.  If one relaxes the CP invariance of $L_Y$,
the phases of $V_R$ become independent of $V_L$ and generically this
lower bound becomes much stronger. Hence the mass of the
flavor-changing Higgs must be determined by $v_R$ rather than $k_+$.
The CP-violating complex phase $\alpha $ enters the quark mass matrix
through the Yukawa interactions in \eq{yuk}. If $L_Y$ in \eq{yuk} is
chosen to conserve CP, the CKM CP violation stems solely from $\alpha
\neq 0$. This case requires that $k_1 k_2 (\sin \alpha)/k_-^2 \approx
m_b/m_t$ \cite{cpwrong,bf,ckmcp}.  In \cite{cpwrong,bf,ckmcp} this has
been achieved by choosing $k_2/k_1 \leq {\cal O} (m_b/m_t)$ and
$\alpha= {\cal O} (1)$. 

We next decompose the Higgs fields into real and imaginary parts:
$\phi_1^0=(\phi_1^{0r} + i \phi_1^{0i} + k_{1})/\sqrt{2}$,
$\phi_2^0=(\phi_2^{0r} + i \phi_2^{0i} + k_{2})\exp(i
\alpha)/\sqrt{2}$ and analogously $\Delta_{L}^0$ and $\Delta_{R}^0$.
\out{
$\Delta_{L}^0 = (\Delta_{L}^{0r} + i
\Delta_{L}^{0i}+ v_{L})\exp(i \delta)/\sqrt{2}$ and $\Delta_{R}^0 =
(\Delta_{R}^{0r} + i \Delta_{R}^{0i}+ v_{R})/\sqrt{2}$.
}
$V$ in \rfn{V} is minimized by solving the equations
\begin{eqnarray}
  \frac{\partial V}{\partial \phi_1^{0r}}= \frac{\partial V}{\partial
    \phi_2^{0r}}= \frac{\partial V}{\partial \phi_2^{0i}}= \frac{\partial
    V}{\partial \Delta_R^{0r}}= \frac{\partial V}{\partial \Delta_L^{0r}}=
  \frac{\partial V}{\partial \Delta_L^{0i}} =0 .\!
  \label{mineq}
\end{eqnarray}
In general \eq{mineq} expresses six chosen parameters of $V$ in terms
of the remaining parameters and $k_1$,$k_2$,$\alpha$,$v_R$,$v_L$ and
$\theta$. Yet there is an important exception: if the parameters in
$V$ are such that CP remains unbroken, the complex phases $\alpha$
and $\theta$ are zero and $V$ is quadratic in $\phi_{1,2}^{0i}$. 
The rank of \eq{mineq} then collapses to 4 allowing us to solve for
only 4 parameters in terms of $k_1$,$k_2$,$v_R$ and $v_L$.
\out{Put conversely, the coupling constants have to fulfill these
relations to generate the required hierarchy of the VEVs.} 
For generic choices of the parameters in $V$ the polynomial equations
in \eq{mineq} will not generate the desired gauge hierarchy $v_R \gg
k_+$. This relation must be encoded in $V$: either some ratio of
dimensionful parameters $\mu_i^2$ or some coupling or a combination of
couplings must be chosen small to define the ratio $k_+^2/v_R^2$.
From this consideration it is clear that fine-tuning is unavoidable.
Parameters fine-tuned to small values must be protected by an
approximate symmetry, otherwise the corresponding solution becomes
unstable under radiative corrections.

Our strategy is to expand the solutions of \eq{mineq} and the Higgs
mass matrices in terms of $\e = \max \{ k_+/v_R, v_L/k_+ \}$. Thereby
we determine the Higgs spectrum in the decoupling limit. We will see
that there are different possibilities to generate the gauge
hierarchy: depending on which parameter is chosen to be of order
$\e^2$, different low energy models emerge in the decoupling limit.
\out{
In
particular the ratio of $\mu_3^2$ and $\rho_1$ will define $v_R$ since
they are quadratic and quartic in $\delta_r^0$. Further the
$\mu_{1,2}^2$ and $\alpha_{1,2,3}$ terms have to cancel up to the weak
scale to generate the proper $k_{1,2}^2$. Hence if we solve
(\ref{mineq}) for $\mu_{1,2,3}$ and $\beta_1$ we find for the leading
$v_R^2$ part
}
We first use the derivatives with respect to $\phi_{1,2}^{0r}$ and
$\Delta_R^{0r}$ in \eq{mineq} to find
\begin{equation}
\!\!  \frac{\mu_1^2}{v_R^2} \! \simeq \! 
  \frac{\alpha_1}{2} \!\! - \! \frac{\alpha_3 k_2^2}{2 k_-^2},  \,\,\,\, 
  \frac{\mu_2^2}{v_R^2} \simeq \frac{\alpha_2}{2} \! 
       + \! \frac{\alpha_3 k_1 k_2}{4 k_-^2 \cos \alpha}, \,\,\,\,
  \frac{\mu_3^2}{v_R^2} \!\simeq \! \rho_1. \!\! \label{minmu}
\end{equation}
These relations are valid in all scenarios discussed in the following.
Here and in the following ``$\simeq$'' means ``equal up to corrections of
order $\e^2$.''
\out{
Here it seems that the hierarchy problem is not present since there is
no fine-tuning in the dimensionful coupling constants. But as we will
see below other parameters have to be small to generate the hierarchy
in the VEVs and thereby imply the extra light Higgs. In particular
either $\alpha$ or $\alpha_3$ have to be small. Together with
(\ref{minmu}) this implies fine-tuning relations between
$\alpha_{1,2,3}$ and $\mu_{1,2}$, while deviations from these
relations define the electroweak scale. To avoid this fine-tuning one
could choose $v_R^2 \alpha_{1,2,3}$ and $\mu_{1,2}$ to be of weak
scale order. Nevertheless the hierarchy problem would still be present
and a small $\alpha_3$ leads, as we will see, to a flavor-violating
light second Higgs doublet.
}
Next we calculate the mass matrices from the second derivatives of $V$
with respect to the Higgs fields and insert the results for the six
parameters found from \eq{minmu}. This step gives us a $2\times 2$
mass matrix for the doubly charged Higgs fields, a $4\times 4$ matrix
for the singly charged Higgses and an $8\times 8$ matrix for the
neutral ones. The latter two contain 2 zero eigenvalues each
corresponding to the pseudo-Goldstone modes eaten by the left- and
right-handed vector bosons. A pivotal role for the mass spectrum is
played by the term involving $\alpha_3$ in $V$, it is the only term
which generates a mass splitting of order $v_R^2$ between the
bidoublet fields:
\begin{equation}
  \alpha_3 \mrm{Tr}( \phi^\dagger\phi \Delta_R \Delta_R^\dagger )
  \rightarrow 
  \alpha_3 |v_R|^2 ( |\phi^+_2|^2+|\phi^0_2|^2 )/2
  . \label{replacepot}
\end{equation}
{ Here the components of $\phi_\pm =(\phi_\pm^0, \phi_\pm^+)$ are
  defined in \eq{defpm}.}
Up to corrections of order $\e^2$ the fields of $\phi_-$ become
components of the Goldstones eaten by $W_L$ and $Z_L$ and of light
Higgs particles, whose masses are of order $k_+$ or smaller.  Hence if
$\alpha_3={\cal O}(1)$, the bidoublet components in $\phi_+$,
acquire masses of order $v_R$.
However, we will be frequently lead to scenarios with $\alpha_3={\cal
  O} (\e^2)$. Then all entries for the bidoublet mass matrices are at
most of order $k_+^2$. From \eq{V} one easily verifies that the terms
which mix bidoublet and triplet fields are ${\cal O} (v_R k_+)$ or
smaller.  This form of the mass matrices implies that the neutral
(charged) Higgs sector has at least four (two) physical Higgs masses
of order $k_+$. Up to terms of order $\e$ the corresponding mass
eigenstates are bidoublet fields. That is, for $\alpha_3={\cal O}
(\e^2)$ one encounters a two-Higgs doublet model (2HDM) in the
decoupling limit $v_R \to \infty$. In view of the flavor-changing
couplings in \eq{nondiaghiggs} such scenarios are unacceptable.

\boldmath
\mysection{Scenarios with $v_L=0$} 
\unboldmath
We will study the $v_L=0$ scenario here for two reasons: first,
it has been used extensively in the literature \cite{deshpande,dgz,dr},
and second, most of the characteristic features of the general case can
be studied from this simplified case.  In addition to \eq{minmu} the
minimization conditions in \eq{mineq} give:
\begin{eqnarray}
  \alpha_3 & = & 
 4 \tilde \lambda \frac{k_-^2}{v_R^2},\quad
\beta_1 = - 2 \beta_3 \frac{k_2}{k_1} \cos \alpha,\quad
\beta_2 = \beta_3 \frac{k_2^2}{k_1^2},  \label{al3} 
\end{eqnarray}
where $\tilde \lambda= 2 \lambda_2- \lambda_3$.  It must be clear that
the scenario with $v_L=0$ is singular, because the six equations in
\eq{mineq} involve only 4 parameters $k_1$,$k_2$,$\alpha$ and $v_R$.
After eliminating $k_1$,$k_2$,$\alpha$ and $v_R$ from \eq{al3} 
one finds two relations between the coefficients of $V$. In
\cite{deshpande,dgz,dr} this problem has been circumvented by choosing
$\beta_{1,2,3}=0$.  Since there is no suitable symmetry, this scenario
cannot be stabilized after renormalization. In particular the desired
Majorana couplings induce non-zero $\beta_i$-terms at the loop level.
\out{ Here we see that the ratio of $\rho_1$ and $\mu_1$ defines
  $v_R$. $\alpha_3$ has to be small to generate the weak scale and
  $\mu_{1,2}$ have to cancel $\alpha_{1,2}$, where their deviation
  defines $k_1/k_2$ and $\alpha$.  Now the VEVs are fixed and
  $\beta_{1,2}$ have to be chosen appropriately to stabilize the
  $v_L=0$ scenario.  }
After eliminating the $\mu_i$'s with \eq{minmu} we find the physical 
eigenvalues of the singly charged mass matrix as
\begin{eqnarray}
  M_1^{+2} &\simeq & \frac{\alpha_3}{2} v_R^2 \frac{k_+^2}{k_-^2}
  , \quad
  M_2^{+2} \, \simeq \, \frac{v_R^2}{2} (\rho_3- 2 \rho_1) 
  . \label{chm}
\end{eqnarray}
Next we use \eq{al3} to eliminate $\alpha_3$.  Since $\alpha_3$ is of
order $\e^2$, this immediately implies that $M_1^+$ in \eq{chm} is of
order $k_+$, not of order $v_R$. 
{ (The smallness of $\alpha_3$ can be motivated by an approximate 
discrete symmetry: $V$ is invariant under $C \equiv C_\phi \circ
C_\Delta $ (defined in \eq{c}). Demanding invariance of $V$ under 
$C_\phi\times C_\Delta $ implies $\alpha_3=0$ (and $\beta_2=\beta_3$).)}  
\out{This example demonstrates that it is necessary to use \emph{all}\ 
  constraints from the minimization conditions.}
The smallness of $M_1^+$ simply reflects our finding that $\alpha_3 =
{\cal O} (\e^2)$ leads to a 2HDM in the decoupling limit, as discussed
after \eq{replacepot}.  The peculiar result for $\alpha_3$ stems from
$\partial V/\partial \phi_2^{0i}$ in \eq{minmu}.  
{ The mechanism here is the following: first 
the heavy scale} $v_R$ is defined by the size of the
dimensionful parameters $\mu_i$ in $V$.  { Then the} first equation in
\eq{al3} tells us that in the chosen scenario (with $v_L=0$ and
$\alpha\neq 0$) the electroweak scale is defined by $k_+^2={\cal
  O}(\alpha_3 v_R^2)$. Choosing $\alpha_3
={\cal O} (1)$ would fail to produce the desired gauge hierarchy $k_+
\ll v_R$.
{ The important lesson is that the same small parameter $\alpha_3$,
which defines the ratio $k_+/v_R={\cal O} (\e)$ through \eq{al3},
enters the physical Higgs masses in \eq{chm}. Only after eliminating
$\alpha_3$ via the minimization conditions \eq{al3} the smallness of
$M_1^+ = {\cal O} (k_+)$ becomes transparent.  From the discussion
preceding \eq{replacepot} we conclude that in the limit $v_R \to
\infty$ the bidoublet does not decouple. What phenomenologically
matters, is of course the masses of the FCNC Higgses. We have
calculated the $8\times 8$ neutral mass matrix and have indeed
verified that there are FCNC Higgs masses of order $\alpha_3 v_R^2$ (or
explicitly of order $k_+^2$). Even if $\tilde \lambda$ in \eq{al3}
is stretched to the largest values compatible with perturbation
theory, the FCNC Higgs masses are way too small to comply with the
precision data from flavor physics.}
\out{
Interestingly, one can find a discrete symmetry protecting
$\alpha_3\approx0$: if one demands invariance of $V$ under $C_{\phi}$
and $C_{\Delta}$ (defined in \eq{c}) separately, the $\alpha_3$ term
in $V$ is forbidden and $\beta_2=\beta_3$.} 
In addition to the unacceptable light FCNC Higgs a fine-tuning problem
emerges in \eq{minmu}: the terms involving $\alpha_3$ are now
sub-leading in $\e$, the dependence on $\alpha$ is lost and the three
parameters $\mu_{1,2,3}$ only depend on $v_R$, up to ${\cal O} (\e^2)$
corrections.  The last two equations in \eq{minmu} now require that
$\mu_3^2/\mu_i^2 \simeq 2\rho_1/\alpha_i$ for $i=1,2$.

A qualitatively new scenario, which has been studied extensively in
\cite{deshpande,dgz,dr}, is obtained, if one chooses $\alpha=0$. Now
there is no SCPV and the rank of \eq{mineq} collapses to four, because
all equations are real. In particular the first equation in \eq{al3}
is now absent and no restrictions on $\alpha_3$ occur! $\alpha_3={\cal
O} (1)$ is now possible, and for this choice we can arrange for a
SM-like Higgs spectrum in the decoupling limit.
 
We demonstrate the impact of $\alpha$ on the mass spectrum for the
case $\beta_i=0$, which can be nicely seen from the mass term of
$\phi_2^{0i}=\imag \phi_2^0$.  After using \eq{minmu} one finds
\begin{eqnarray}  
&& \!\!\!\!
\phi_2^{0i} \lt. \frac{\partial V}{\partial \phi_2^{0i}} \rt|_{\phi_2^{0i}=0}
+ \lt. \frac{\phi_2^{0i\,2}}{2} \frac{\partial^2 V}{\partial
  \phi_2^{0i\,2}} \rt|_{\phi_2^{0i}=0}
 \, = \, 
\nn
&&\qquad 
\phi_2^{0i} k_2 \sin \alpha \lt[ 
   \frac{\alpha_3}{2} v_R^2 - 2  ( 2 \lambda_2 -\lambda_3
   ) k_-^2 \rt] \nn 
&&\qquad + 
   \phi_2^{0i\, 2} \lt[ 
    \frac{\alpha_3}{4} v_R^2 + k_2^2 \sin^2 \alpha - 
        k_-^2 (2 \lambda_2 -\lambda_3)     \rt] 
 \label{mph2}.
\end{eqnarray}  
For $\alpha\neq 0$ the fourth minimization equation (w.r.t.\ 
$\phi_2^{0i}$) enforces the linear term to vanish yielding the
condition for $\alpha_3$ in \eq{al3}. Then the low energy model is a
2HDM. If $\alpha =0$, however, the linear term is zero automatically,
and $\alpha_3$ can be of order 1, so that both charged Higgs boson
masses in \eq{chm} are naturally of order $v_R^2$. Put conversely,
$\alpha_3 = {\cal O}(1)$ implies $\alpha=0$ and a heavy second Higgs
multiplet, while SCPV requires a small $\alpha_3$ and thereby implies
a second light doublet. Hence the CP properties of the vacuum state
are connected to the Higgs spectrum in the decoupling limit. We
further stress that the scenario with $\alpha=0$ cannot be obtained by
taking the limit $\alpha \to 0$ from the general case.  For $\alpha=0$
and $\alpha_3={\cal O}(1)$ we find a SM-like Higgs spectrum for $v_R
\to \infty$ in agreement with \cite{deshpande,dgz,dr}. However, we
again face a fine-tuning problem, because in \eq{minmu} the three
equations only involve two parameters $v_R$ and $k_2/k_1$. By
eliminating $v_R$ and $k_2/k_1$ from \eq{minmu} one easily finds a
relation between $\mu_{1,2,3}$, $\alpha_{1,2,3}$ and $\rho_1$ which
cannot be justified by any symmetry.
\out{
Even if one found a symmetry enforcing $\mu_{1,2}^2/v_R^2$ 
and $\alpha_{1,2}$ (which multiply $\mrm{Tr}(\phi^\dagger \phi )$ and
$\mrm{Tr}(\tilde{\phi}^\dagger \phi )$) to be ${\cal O} ( \e^2)$, \eq{minmu}
would require $\alpha_3={\cal O}(\e^2)$ and thus lead to a 2HDM.}
\out{Moreover the electroweak scale $k_+$ would be defined by the
  ${\cal O} (\e^2)$ corrections to \eq{minmu}.}

\mysection{General Scenarios}
Next we will study general scenarios with $v_L\neq 0$.  We solve the
minimization conditions for $\mu_1^2$, $\mu_2^2$, $\rho_1,$ $\rho_3$, 
$\beta_1,$ and $\beta_2$.  The solutions for $\mu_1$, $\mu_2$ and 
$\rho_1$ can be found in \eq{minmu}.  
For $\rho_3$, $\beta_1$ and $\beta_2$ we find:  
\begin{eqnarray}
\rho_3  &=&  
  \frac{\alpha_3 \, \sin^2 \alpha \, 
    k_1^2 k_2^2}{v_L^2 \! \lt[ k_1^2 \sin^2 \theta  - 
  k_2^2 \lt(\sin^2 \alpha + \sin^2 ( \alpha + \theta)  \rt) \rt]  } 
          + {\cal O} (\e^0), \nn
          \beta_1 &\simeq&   \rho_3 \frac{v_L v_R}{k_1 k_2} 
                \frac{\sin \theta}{\sin \alpha}, \qquad
  \beta_2 \; \simeq \; - \beta_1 \frac{k_2}{k_1} 
       \frac{ \sin (\alpha+\theta) }{\sin \theta} .
\label{surp}
\end{eqnarray}
If both $\alpha$ and $f\equiv k_1 k_2/(v_L v_R)$ are of order 1, one
faces non-perturbatively large couplings
$\beta_1$,$\beta_2$,$\rho_3={\cal O} (\e^{-2})$, unless $\alpha_3 =
{\cal O} (\e^2)$. \eq{surp} can be viewed as a see-saw relation
between $1/\alpha_3$ and $\beta_1,\rho_3$. The pivot is proportional
to $\sin \alpha$, so that the effect vanishes in the absence of the
CP-violating phase.  Its origin can be traced back to the terms in $V$
which involve $\phi_2^{0i}$, as in (\ref{mph2}).  Hence we confirm the
findings of \cite{deshpande,lrh} that $\alpha={\cal O}(1)$ and $k_1
\sim k_2$ enforces the smallness of parameters in the potential.
However, this feature also appears in all other scenarios, as we
discussed above. Hence an exhaustive investigation of the Higgs sector
requires to consider all these possibilities that CP phases, ratios of
VEVs and some of the Higgs couplings scale with certain powers of
$\e$. We have performed a complete analysis of all possible scenarios
and discuss the generic cases here.

\mysubsection{Small $k_2 \sin \alpha$} 
If $\rho_3$ and $\alpha_3$ are of order 1, \eq{surp} shows that
$(k_2/k_+) \sin \alpha$ is ${\cal O} (\e^2)$ or smaller.
For $(k_2/k_+) \sin \alpha = {\cal O} (\e^2)$ the triplet
phase $\theta$ must be of order 1. Smaller values of $\theta$ are
correlated with even tinier values of $\alpha$.
\out{
Here \eq{surp} might be satisfied by a) $\alpha={\cal O} (\e^2)$, b)
$k_{2(1)}={\cal O} (\e^2)$ or if both c) are ${\cal O} (\e)$.
}
Interestingly, one now finds the relation:
\begin{equation}
2\rho_3- \rho_1 = {\cal O} (\e^2). \label{eq:nocpfine-tune}
\end{equation}
That is, in this scenario the electroweak scale is defined by
$k_+^2={\cal O} ((2\rho_3- \rho_1) v_R^2)$, while $v_R^2$ is as
usual defined by $\mu_3^2/\rho_1$ through \eq{minmu}. In any GUT
scenario in which $\Delta_L$ and $\Delta_R$ belong to the same GUT
multiplet, $2\rho_3- \rho_1$ vanishes exactly above the GUT scale.
Below the GUT scale $2\rho_3- \rho_1$ acquires a small non-zero value from
radiative corrections. Such a situation occurs for example in SO(10)
models \cite{dr}. The responsibility of the GUT symmetry for the gauge
hierarchy $k_+ \ll v_R$ is certainly very interesting. However, the
small parameter $2\rho_3- \rho_1$ also proliferates into the mass
matrices, the leading terms for the singly charged masses are as in
\eq{chm}.  While the doublet with the FCNC Higgs now becomes heavy, we
instead face an extra light Higgs triplet, whose components are
dominantly $\Delta_L$ fields.  The explicit calculation of the neutral
mass matrix, however, shows that the triplet masses are of order $v_L$
or smaller. Since at least one of the triplet fields couples to the
Z-boson, this scenario would have been discovered at LEP-I. These
findings also hold for the case that $k_2 \sin \alpha$ is exactly zero
with $\theta \neq 0$.

If both $(k_2/k_+)\sin \alpha$ and $\alpha_3$ are ${\cal
  O}(\e)$, we find three light neutral Higgses: one is SM-like one and
two are mixtures of the $\delta_L^0$ and the flavor-violating $\phi_+^0$.
This scenario interpolates between the previous and the 
following\nolinebreak~ one.

\mysubsection{Large $k_2 \sin \alpha$}
For $(k_2/k_+)\sin \alpha={\cal O}(1)$ the minimization conditions in
\eq{surp} require $\alpha_3 ={\cal O} (\e^2)$, thus $\alpha_3$ defines
the gauge hierarchy here. This scenario agrees qualitatively with the
one discussed above for $v_L=0$.  As stated previously the smallness
of $\alpha_3$ leads to a 2HDM in the decoupling
limit, with unacceptable flavor-violating neutral couplings in
(\ref{nondiaghiggs}). We have also calculated the renormalization
group flow of the flavor-changing neutral Higgs couplings in order to
rule out a possible suppression mechanism of these couplings. 

\mysubsection{No SCPV}
For $\alpha=\theta=0$ the minimization conditions with respect to
$\Delta_{L}^{0i}$ and $\phi_{2}^{0i}$, which implied the smallness of
$\alpha_3$ or $2\rho_3-\rho_1$, vanish and we are just left with 4
minimization conditions. Now we can arrange the non-standard Higgses
to decouple for $v_R \to \infty$. Also this scenario agrees
qualitatively with the zero phase case with $v_L=0$: (\ref{minmu})
implies a severe fine-tuning problem, after eliminating $v_R$ and
$k_2/k_1$ from \eq{minmu} the parameters $\mu_{1,2,3}$,
$\alpha_{1,2,3}$ and $\rho_1$ combine to a relation requiring that
parameters of order 1 cancel up to terms of order $\e^2$. This
phenomenon is at the heart of the gauge hierarchy problem. One usually
addresses it by choosing the mass parameters $\mu_{1,2}$ of the fields
which break the electroweak symmetry to be of order $k_+$. That is,
the gauge hierarchy $k_+ \ll v_R$ is put into $V$ by hand. In our case
this solution would also require to choose $\alpha_i= {\cal O}(\e^2)$,
i=1,2,3, (see \eq{minmu}) leading again to a 2HDM.

\boldmath 
\mysection{ CP violation in $V$ at a high scale} 
\unboldmath 
{ The essential prerequisite for the connection between the CP
phases and the Higgs spectrum found above is the spontaneous breakdown
of CP at the \emph{electroweak}\ scale. Within the MLRM a new
situation occurs, if one allows for explicit CP violation in $V$. In
this case} only the term involving $\alpha_2$ in \eq{V} changes (see
Eq.~(A.2) of \cite{deshpande}) and $\alpha_2$ may be complex. To order
$\e^0$ this case can be mapped on the discussed one by rescaling
$\phi_2^0 \to \phi_2^0 { e^{-i \varphi} }$ with $ \varphi=\arg
(\alpha_2 v_R^2/2 -\mu_2^2)$.  Hence the previous findings on the
Higgs spectrum in the decoupling limit remain valid with the
replacement $\alpha \to \alpha + { \varphi} $. 
{ Now the SM-like spectrum occurs for $\alpha=-\varphi $. Going
beyond the MLRM, one can add extra singlet Higgs fields with large VEVs of
order $v_R$ \cite{bl}. Then spontaneous CP phases can appear between
different VEVs of order $v_R$, i.e.\ CP is broken well above the
electroweak scale. This case can be mapped on the case with explicit
CP violation, with $\varphi$ now being a calculable function of the
new CP phases. Phenomenological studies of CKM CP violation in these
models requires a renormalization group analysis of the Yukawa sector 
to account for the large logarithm $\ln (v_R/k_+)$.}

\mysection{Conclusions}
We have determined the general Higgs potential of the MLRM in the
decoupling limit $v_R \to \infty$ and established a strict connection
between the CP properties of the vacuum state and the Higgs spectrum:
if any of the CP phases $\alpha + { \varphi}$ or $\theta$ is
nonzero, the low-energy model always differs from the SM. Either a
2HDM with flavor-changing Higgses or a model with additional light
triplet fields emerges, both of which are phenomenologically ruled
out. { The appearance of the triplet Higgs model in the decoupling
limit for small but non-zero $\alpha$ has not been discussed in
previous analyses.}  In particular { papers which simultaneously
assume a large $\alpha+\varphi$ and a SM-like Higgs spectrum} are not
correct.  { An important difference to previous analyses of the
Higgs sector is that our results hold generally, even for fine-tuned
parameters. We found that fine-tuning arguments do not discriminate
between scenarios with and without spontaneous CP phases. Fine-tuning
is unavoidable for the Higgs potential to produce the gauge hierarchy
$v_R \gg k_+$.}  To obtain the SM in the decoupling limit in the case
$\alpha{ +\varphi}=\theta=0$, a fine-tuning condition between
$\mu_{1,2,3}$, $\alpha_{1,2,3}$ and $\rho_1$ { implied by
\eq{minmu}} must be fulfilled. This condition requires parameters of
order $v_R^2$ to cancel up to terms of order $k_+^2$ and cannot be
justified by an approximate symmetry.  { The main conclusion of our
paper is that the MLRM does not allow for spontaneous CP violation at the
electroweak scale. Adding extra singlet fields does not change this
conclusion, but opens the possibility for spontaneous CP violation at
the high scale $v_R$ parametrized by a new phase $\varphi$. Decoupling of
the non-standard Higgses then requires $\alpha = - \varphi$.}

This work is supported by BMBF under contract No.~05HT1WOA3, by the
EU under the TMR contract No.~HPMF-CT-2000-00460 and by ESF grant No.
3832.  Fermilab is operated by URA under DOE contract
No.~DE-AC02-76CH03000.
 
\vspace{-3ex}
\bibliographystyle{unsrt}

\end{document}